\documentclass{jfm}

\usepackage{graphicx}
\usepackage{natbib}
\usepackage{epstopdf, epsfig}
\usepackage{amsmath}
\usepackage{color}
\usepackage{subfigure}
\usepackage{dashrule}

\def\includefigs{\let\ifincfigs=\iftrue}
\def\noincludefigs{\let\ifincfigs=\iffalse}
\includefigs

\newbox\epsfvertlab
\newbox\epsfhorlab
\newbox\epsffiglab

\newdimen\epsfvlabsize
\newdimen\scott

\def\setvlabel#1{\setbox\epsfvertlab=\vbox{\hbox{#1}}}%
\def\sethlabel#1{\setbox\epsfhorlab=\vbox{\hbox{#1}}}%
\def\figlab#1 #2 #3{\setbox\epsffiglab=\vbox to 0pt{%
\ifvoid\epsffiglab\else\box\epsffiglab\fi\vss\hbox to 0pt{\raise #2 \hbox{\hskip #1 #3}\hss}}}

\newdimen\fighor
\newdimen\figver
\newbox\rotbox
\long\def\lrlap#1{\hbox to 0pt{#1\hss}}
\long\def\verttex#1#2#3{{\fighor = #1\figver = #2\vbox to \figver{\vss%
\hbox to \fighor{\hfill\hsize=\fighor%
\lrlap{\rotstart{-90 rotate}\vbox to \fighor{#3\vfil}\rotfinish}}}}}

\def\dvipsvspec#1{\special{ps:#1}}
\def\dvipsrotstart#1{\dvipsvspec{gsave currentpoint currentpoint translate
   #1 neg exch neg exch translate}}
\def\dvipsrotfinish{\dvipsvspec{currentpoint grestore moveto}}

\def\rotstart#1{\dvipsrotstart{#1}}
\def\rotfinish{\dvipsrotfinish}

\def\epsfsetlab{%
\ifvoid\epsfvertlab%
\else%
\verttex{\epsfvlabsize}{\epsfysize}%
{\hbox to \epsfysize{\hss\box\epsfvertlab\hss}}%
\fi%
\ifvoid\epsfhorlab%
\else%
\scott=\epsfxsize%
\advance\scott by \epsfvlabsize%
\rlap{\vtop{\hrule height0pt\hbox to \scott{\hss\box\epsfhorlab\hss}}}%
\fi%
}

\def\epsfsetover{\ifvoid\epsffiglab\else\box\epsffiglab\fi}

\newread\epsffilein    
\newif\ifepsffileok    
\newif\ifepsfbbfound   
\newif\ifepsfverbose   
\newdimen\epsfxsize    
\newdimen\epsfysize    
\newdimen\epsftsize    
\newdimen\epsfrsize    
\newdimen\epsftmp      
\newdimen\pspoints     
\def\epsfbox#1{
   \ifvoid\epsfvertlab%
   \else\epsfvlabsize=\ht\epsfvertlab \advance\epsfvlabsize by \dp\epsfvertlab\fi%
   \leavevmode\global\def\epsfllx{72}\global\def\epsflly{72}%
   \global\def\epsfurx{540}\global\def\epsfury{720}%
   \def\lbracket{[}\def\testit{#1}\ifx\testit\lbracket
   \let\next=\epsfgetlitbb\else\let\next=\epsfnormal\fi\next{#1}}%
\def\epsfgetlitbb#1#2 #3 #4 #5]#6{\epsfgrab #2 #3 #4 #5 .\\%
   \epsfsetgraph{#6}}%
\def\epsfnormal#1{\epsfgetbb{#1}\epsfsetgraph{#1}}%
\def\epsfgetbb#1{%
\openin\epsffilein=#1
\ifeof\epsffilein\errmessage{I couldn't open #1, will ignore it}\else
   {\epsffileoktrue \chardef\other=12
    \def\do##1{\catcode`##1=\other}\dospecials \catcode`\ =10
    \loop
       \read\epsffilein to \epsffileline
       \ifeof\epsffilein\epsffileokfalse\else
          \expandafter\epsfaux\epsffileline:. \\%
       \fi
   \ifepsffileok\repeat
   \ifepsfbbfound\else
    \ifepsfverbose\message{No bounding box comment in #1; using defaults}\fi\fi
   }\closein\epsffilein\fi}%
\def\epsfsetgraph#1{%
   \epsfrsize=\epsfury\pspoints
   \advance\epsfrsize by-\epsflly\pspoints
   \epsftsize=\epsfurx\pspoints
   \advance\epsftsize by-\epsfllx\pspoints
   \epsfxsize\epsfsize\epsftsize\epsfrsize
   \ifnum\epsfxsize=0 \ifnum\epsfysize=0
      \epsfxsize=\epsftsize \epsfysize=\epsfrsize
     \else\epsftmp=\epsftsize \divide\epsftmp\epsfrsize
       \epsfxsize=\epsfysize \multiply\epsfxsize\epsftmp
       \multiply\epsftmp\epsfrsize \advance\epsftsize-\epsftmp
       \epsftmp=\epsfysize
       \loop \advance\epsftsize\epsftsize \divide\epsftmp 2
       \ifnum\epsftmp>0
          \ifnum\epsftsize<\epsfrsize\else
             \advance\epsftsize-\epsfrsize \advance\epsfxsize\epsftmp \fi
       \repeat
     \fi
   \else\epsftmp=\epsfrsize \divide\epsftmp\epsftsize
     \epsfysize=\epsfxsize \multiply\epsfysize\epsftmp   
     \multiply\epsftmp\epsftsize \advance\epsfrsize-\epsftmp
     \epsftmp=\epsfxsize
     \loop \advance\epsfrsize\epsfrsize \divide\epsftmp 2
     \ifnum\epsftmp>0
        \ifnum\epsfrsize<\epsftsize\else
           \advance\epsfrsize-\epsftsize \advance\epsfysize\epsftmp \fi
     \repeat     
   \fi
   \ifepsfverbose\message{#1: width=\the\epsfxsize, height=\the\epsfysize}\fi
   \epsftmp=10\epsfxsize \divide\epsftmp\pspoints
   \epsfsetlab%
   \ifincfigs%
     \vbox to\epsfysize{\vfil\hbox to\epsfxsize{%
        \includegraphics{#1}%
        \epsfsetover\hfil}}%
   \else%
     \epsfsetover%
     \vbox to\epsfysize{\hrule\vss\hbox to\epsfxsize{\vrule height
                        \epsfysize\hfil\vrule}\vss\hrule}%
   \fi%
\epsfxsize=0pt\epsfysize=0pt}%

{\catcode`\%=12 \global\let\epsfpercent=
\long\def\epsfaux#1#2:#3\\{\ifx#1\epsfpercent
   \def\testit{#2}\ifx\testit\epsfbblit
      \epsfgrab #3 . . . \\%
      \epsffileokfalse
      \global\epsfbbfoundtrue
   \fi\else\ifx#1\par\else\epsffileokfalse\fi\fi}%
\def\epsfgrab #1 #2 #3 #4 #5\\{%
   \global\def\epsfllx{#1}\ifx\epsfllx\empty
      \epsfgrab #2 #3 #4 #5 .\\\else
   \global\def\epsflly{#2}%
   \global\def\epsfurx{#3}\global\def\epsfury{#4}\fi}%
\def\epsfsize#1#2{\epsfxsize}
\def\ifspace{\ifcat\issp.\else~\fi}
\def\tspace{\futurelet\issp\ifspace}
\def\a{({\it a\kern 1pt})\tspace}
\def\b{({\it b\kern 1pt})\tspace}
\def\c{({\it c\kern 1pt})\tspace}
\def\d{({\it d\kern 1pt})\tspace}
\def\e{({\it e\kern 1pt})\tspace}
\def\f{({\it f\kern 1pt})\tspace}
\def\g{({\it g\kern 1pt})\tspace}
\def\h{({\it h\kern 1pt})\tspace}
\def\i{({\it i\kern 1pt})\tspace}
\def\j{({\it j\kern 1pt})\tspace}
\def\abc#1{({\it #1\kern 1pt})\tspace}

\newcount\ndots
\def\drawline#1#2{\raise 2.5pt\vbox{\hrule width #1pt height #2pt}}

\def\trian{\raise 1.25pt\hbox{$\scriptscriptstyle\triangle$}\nobreak\ }
\def\solidtrian{\raise 1.25pt
\hbox to 3bp{
\hfill}\nobreak\ }
\def\dsolidtrian{\raise 1.25pt
\hbox to 3bp{
\hfill}\nobreak\ }
\def\soliddiamond{\raise 1.25pt
\hbox to 4bp{
\hfill}\nobreak\ }

\def\square{${\vcenter{\hrule height .4pt 
              \hbox{\vrule width .4pt height 3pt \kern 3pt \vrule width .4pt}
          \hrule height .4pt}}$\nobreak\ }

\def\plus{\raise 1.25pt \hbox{$\scriptscriptstyle +$}\nobreak\ }
\def\x{\raise 1.25pt \hbox{$\scriptscriptstyle \times$}\nobreak\ }
\def\legendtable#1{\vbox{\baselineskip=10pt\tabskip=0pt\let\\=\cr\halign{\hfil##\hskip 3pt&##\hfil\cr#1\crcr}}}
\def\lllegend#1 #2 #3{\figlab {#1} {#2} {\legendtable{#3}}}
\def\lrlegend#1 #2 #3{\figlab {#1} {#2} {\llap{\legendtable{#3}}}}
\def\ullegend#1 #2 #3{\figlab {#1} {#2} {\vtop{\hrule height 0pt\legendtable{#3}}}}
\def\urlegend#1 #2 #3{\figlab {#1} {#2} {\llap{\vtop{\hrule height 0pt\legendtable{#3}}}}}

\newdimen\xorigon
\newdimen\yorigon
\newdimen\scaleval
\newdimen\scaleorigon

\def\setxscale#1 #2 #3 #4 #5 {%
    \xorigon=#1\yorigon=#3%
    \scaleval=#2\advance\scaleval by -\xorigon%
    \tempdimen=#5 pt\advance\tempdimen by -#4pt%
    \divide\tempdimen by 1000%
    \divide\scaleval by \tempdimen%
    \scaleorigon=-#4pt\divide\scaleorigon by 1000%
    \multiply\scaleorigon by \scaleval}
\def\xtickup#1 #2{\tempdimen=#1pt\divide\tempdimen by 1000%
    \multiply\tempdimen by \scaleval\advance\tempdimen by \scaleorigon%
    \advance\tempdimen by \xorigon%
    \figlab {\tempdimen} {\yorigon} {\vbox {\hbox to 0pt{\hss #2\hss}%
        \baselineskip=8pt\lineskiplimit=-5pt%
        \hbox to 0pt{\hss \vrule height 3pt\hss}}}}
\def\xtickdown#1 #2{\tempdimen=#1pt\divide\tempdimen by 1000%
    \multiply\tempdimen by \scaleval\advance\tempdimen by \scaleorigon%
    \advance\tempdimen by \xorigon%
    \figlab {\tempdimen} {\yorigon} {\vbox to 0pt {\hbox to 0pt{\hss \vrule height 3pt\hss}%
        \nointerlineskip\vskip 3pt%
        \hbox to 0pt{\hss #2\hss}\vss}}}
\def\nofig#1#2{\leavevmode{\vbox {\hrule \hbox to #1{\vrule height #2 \hfill \vrule} \hrule}} }

\graphicspath{{./figures/}}
\DeclareGraphicsExtensions{.eps}

\newcommand{\bea}{\begin{equation}\begin{aligned}}
\newcommand{\eea}{\end{aligned}\end{equation}}

\shorttitle{Axisymmetric boundary layers}
\shortauthor{P. Kumar and K. Mahesh}

\title{Analysis of axisymmetric boundary layers}

\author{Praveen Kumar\aff{1}
\and Krishnan Mahesh\aff{1}
\corresp{\email{kmahesh@umn.edu}}}

\affiliation{\aff{1}Department of Aerospace Engineering and Mechanics, University of Minnesota, Minneapolis, MN 55455, USA}

\begin{document}

\maketitle

\begin{abstract}
Axisymmetric boundary layers are studied using integral analysis of the governing equations for axial flow over a circular cylinder. The analysis includes the effect of pressure gradient and focuses on the effect of transverse curvature on boundary layer parameters such as shape factor ($H$) and skin-friction coefficient ($C_f$), defined as $H = \delta^*/\theta$ and $C_f = \tau_w/(0.5\rho U_e^2)$ respectively, where $\delta^*$ is displacement thickness, $\theta$ is momentum thickness, $\tau_w$ is the shear stress at the wall, $\rho$ is density and $U_e$ is the streamwise velocity at the edge of the boundary layer. Relations are obtained relating the mean wall-normal velocity at the edge of the boundary layer ($V_e$) and $C_f$ to the boundary layer and pressure gradient parameters. The analytical relations reduce to established results for planar boundary layers in the limit of infinite radius of curvature. The relations are used to obtain $C_f$ which shows good agreement with the data reported in the literature. The analytical results are used to discuss different flow regimes of axisymmetric boundary layers in the presence of pressure gradients.  
\end{abstract}

\section{Introduction}

\begin{figure}
\centering{
\includegraphics[width=80mm]{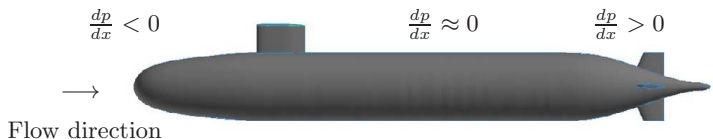}
\put(-240,60){$\frac{dp}{dx} < 0$}
\put(-120,60){$\frac{dp}{dx} \approx 0$}
\put(-50,60){$\frac{dp}{dx} > 0$}
\put(-250,35){$\longrightarrow$}
\put(-270,20){Flow direction}
} 
\caption{Different pressure gradient conditions experienced by the streamwise ($x$) evolving hull boundary layer on a generic submarine hull, AFF8 \citep{grovesgeometry}.}
\label{fig:sub}
\end{figure}

Turbulent boundary layers (TBL) are one of the most studied canonical fluid problems but most past studies are devoted to the flat plate (planar) TBL. A recent review by \cite{TBLreview2011} describes the current understanding and future challenges of wall-bounded flows at high Reynolds number ($Re$). A variety of hydrodynamic engineering applications however, involve axisymmetric TBL, which involve an additional length scale parameter to account for curvature. Several engineering applications have axisymmetric TBL evolving under the influence of pressure gradients due to their geometrical shapes. For example, figure \ref{fig:sub} shows a generic submarine hull \citep{grovesgeometry} along with the streamwise varying pressure gradients experienced by the hull boundary layer.

 The radius based Reynolds number ($Re_a = aU/\nu$, where $U$ is freestream velocity, $\nu$ is kinematic viscosity and $a$ is the radius of cylinder) does not include any effect of wall-shear stress or boundary layer thickness. Therefore, popular non-dimensional parameters to characterize axisymmetric TBL are the ratio of boundary layer thickness to the radius of curvature ($\delta/a$) and the radius of curvature in wall units ($ a^+$). Based on these two parameters, three regimes can be identified \citep{patel}: (i) both  $\delta/a$ and $a^+$ are large, (ii) large $\delta/a$ and small $a^+$ and (iii) small $\delta/a$ and large $a^+$. The first flow regime is observed for axial flow over a long slender cylinder at high $Re$, where large effect of curvature is felt. The second flow regime is realized for axial flow over slender cylinders at low $Re$, where axisymmetric TBL behaves like an axisymmetric wake with an inner layer with strong curvature and low-Re effects. Almost all the experimental studies reported in the literature have focused on the first two regimes \citep[see][]{patel}. The third flow regime is common in applications where the Reynolds number is high but the boundary layer is thin compared to the radius of curvature. Usually, this flow regime is treated as a planar boundary layer where the curvature effects are assumed minimal. Although, there are significant fundamental differences between a planar TBL and a thin axisymmetric TBL at high $Re$, such as increased skin-friction and rapid radial decay in turbulence away from the wall \citep{lueptow1990}.  
 
 One of the earliest analytical investigation of the effect of transverse curvature on skin-friction was conducted by \cite{landweber1949}, who used a $1/7^{th}$-power-law for velocity profile and the Blasius skin-friction law \citep{schlichting} to show that for a given momentum thickness ($\theta$) based Reynolds number ($Re_\theta$), axisymmetric boundary layers have higher skin-friction and lower boundary layer thickness in comparison to planar boundary layers. \cite{seban} analysed the laminar boundary layer for axial flow over a circular cylinder from the governing boundary layer equations and showed that the skin-friction and heat-transfer coefficients for axisymmetric laminar boundary layers are higher than that obtained from the Blasius solution. \cite{kelly} introduced an important correction to their solution, known as the Seban-Bond-Kelly (SBK) solution for zero pressure gradient (ZPG) axisymmetric boundary layers. The SBK solution was extended to the regime of large curvature effect as encountered in axial flow over long thin cylinders by \cite{gl}. \cite{stewartson} provided an asymptotic solution for ZPG laminar axial flow over long thin cylinders. 
 
 Axisymmetric TBL have not received the same attention as planar TBL likely due to the inherent difficulties in keeping the flow perfectly axial and prevent sagging or elastic deformation of the cylinders. The effect of curvature has been the focus of most past studies. \cite{richmond1957} and \cite{yu1958} conducted the first few experimental studies for curvature effects on boundary layers, which was followed by extensive experimental studies \citep{rao1967,cebeci1970,rao1972,chase1972,patel1974,patel1974expt,willmarth1976,luxton1984,lueptow1985,krane2010} showing that the transverse curvature indeed has a significant effect on the overall behaviour of axisymmetric TBL. 

 \cite{afzal} analysed thin axisymmetric TBL at high $Re$ (regime 3 described above) using asymptotic expansions and modified the well-known classical law of the wall for planar TBL to include the effect of curvature. The wall-normal distance in wall units ($y^+$) was modified as,
 \begin{eqnarray}\label{eq:yplus}
 y^+ & = &  a^+ln(1 + y/a) 
 \end{eqnarray}
 where, $a^{+} =   au_{\tau}/{\nu}$ is the radius of curvature in wall-units. Using this modified $y^+$, it was shown that there exists a log layer in the mean velocity profile similar to that found in planar TBL, with same slope but the intercept ($B$) is a weak function of curvature ($B= 5 + 236/a^+$).  It has been shown that $U^+ = a^+ln(1 + y/a)$ is valid in the viscous sublayer region, but the use of $y^+$ from eq. \ref{eq:yplus} instead of the planar $y^+$ in the logarithmic region assumes that transverse curvature affects both the viscous sublayer and log layer identically. 
    
 One of the earliest numerical simulations of axisymmetric boundary layers were performed by \cite{cebeci1970}, who showed higher skin-friction compared to flat plate prediction in both laminar and turbulent regimes. Similar behaviour of skin-friction was observed in numerous subsequent simulations of axisymmetric TBL. Axisymmetric TBL over long thin cylinders have been extensively studied by \cite{tutty} using Reynolds-averaged Navier--Stokes (RANS) and \cite{jordanstats,jordanskin,jordan2014jfe,jordanmodel} using direct numerical simulations (DNS) and large eddy simulations (LES). Jordan used his simulation database to propose simple models for the skin-friction \citep{jordanskin} and the flow field \citep{jordanmodel}. 
 
 None of the studies mentioned so far have considered pressure gradient effects. Experiments by \cite{fernholz1998} and \cite{warnack1998} considered axisymmetric TBL under favourable pressure gradient (FPG) in internal flow.  
 
Boundary layers under adverse pressure gradients (APG) have been studied in the past using asymptotic expansions (See \citet{afzal1983,afzal2008} and references therein). Recently, \cite{wei2016} performed integral analysis of the governing equations for ZPG boundary layers over flat plates and obtained,
\begin{eqnarray}
 \frac {U_e V_e}{u_\tau^2} = H
\label {eq:hflat}
\end{eqnarray}
where $U_e$ and $V_e$ are the mean streamwise and wall-normal velocity at the edge of the boundary layer respectively, $H$ is the shape factor and $u_\tau = \sqrt{\tau_w/\rho}$ is the friction velocity. The analysis was later extended for planar boundary layers under pressure gradient by \cite{wei2017}, which modified eq. \ref{eq:hflat} as,
\begin{eqnarray}
 \frac {U_e V_e}{u_\tau^2} = H + (1 + \delta/\delta^* + H)\beta_{_{RC}}
\label {eq:hflatpress}
\end{eqnarray}
where $\beta_{_{RC}}$ is the Rotta--Clauser pressure gradient parameter \citep{rotta,clauser}, $\delta^*$ is the displacement thickness and $\delta$ is the boundary layer thickness. $\beta_{_{RC}}$ is often used to quantify the strength of APG in boundary layer flows.

The goal of the present work is to analyse the governing equations of axisymmetric boundary layers evolving under the influence of pressure gradient and understand the effect of transverse curvature on the flow. Integral analysis of the governing equations is performed in \S \ref{sec:gov} and the obtained relations are compared to the existing data in \S \ref{sec:valid}. Implications of analytical relations are discussed in \S \ref{sec:dis}. \S \ref{sec:sum} concludes the paper.

\section{Integral analysis of axisymmetric boundary layer}
\label{sec:gov}
The boundary layer approximations for the time-averaged Navier--Stokes equations in cylindrical coordinates yield,
\begin{eqnarray}
\label{eq:bl1}
r\frac{\partial U}{\partial x} + \frac{\partial (rV)}{\partial r} & = & 0, \\
\label{eq:bl2}
rU\frac{\partial U}{\partial x} + rV\frac{\partial U}{\partial r} & = & -\frac{r}{\rho}\frac{dP}{dx} + \frac {\partial {(r \nu \frac {\partial U}{\partial r}})} {\partial r} + \frac {\partial (-r\overline{u'v'})} {\partial r}
\end{eqnarray}
where $U$ and $V$ are mean, and $u'$ and $v'$ are fluctuations in axial and radial velocities respectively. Note that the stress term involving $\partial {(\overline{u'u'}-\overline{v'v'})}/\partial{x}$ has been ignored on the right hand side of eq. \ref{eq:bl2} for the present analysis. This term however, can not be neglected for large magnitude of pressure gradients and boundary layers on the verge of separation. We have not made any assumption on the nature of boundary layer i.e. it can be laminar, transitional or turbulent. This implies that the present analysis holds as long as the governing equations (eqs. \ref{eq:bl1}, \ref{eq:bl2}) are valid.  

For boundary layer under pressure gradient, the mean wall-normal velocity outside the boundary layer ($V_o$) is not constant. Hence, the boundary layer equations are integrated in wall-normal direction from the surface, $r=a$ to a location outside the boundary layer, $r=a + k\delta$ where $a$ is the radius of curvature (cylinder), $k \ge 1$ is a parameter and $\delta$ is the boundary layer thickness. Note that setting $k=1$ makes $V_o = V_e$, which is the mean wall-normal velocity at the edge of the boundary layer. Integration of eqs. \ref{eq:bl1} and \ref{eq:bl2} with the aforementioned limits yield,
\begin{eqnarray}
\label{eq:ibl1}
\int_{a}^{a + k\delta} r\frac{\partial U}{\partial x} dr & = & -\int_{a}^{a + k\delta}\frac{\partial (rV)}{\partial r} dr = -\bigg(rV\bigg)\bigg|_a^{a + k\delta} \nonumber \\
          & = & -(a + k\delta)V_o, \\
\int_{a}^{a + k\delta} rU\frac{\partial U}{\partial x} dr + \int_{a}^{a + k\delta}  rV\frac{\partial U}{\partial r} dr & = & -\int_{a}^{a + k\delta}\frac{r}{\rho}\frac{dp}{dx}dr + \int_{a}^{a + k\delta} \frac {\partial {(r \nu \frac {\partial U}{\partial r}})} {\partial r} dr +  \int_{a}^{a + k\delta} \frac {\partial (-r\overline{u'v'})} {\partial r} dr \nonumber \\
 & = &  -\beta_{_{RC}}\frac{u_\tau^2}{2\delta^*} r^2\bigg|_a^{a + k\delta} + \bigg(r \nu \frac {\partial U}{\partial r}\bigg)\bigg|_a^{a + k\delta} - (r\overline{u'v'})\bigg|_a^{a + k\delta}
 \label{eq:ibl2}
\end{eqnarray}
where $\beta_{_{RC}}$ is defined as,
\begin{eqnarray}
\beta_{_{RC}}  = \frac{\delta^*}{u_\tau^2} \frac{1}{\rho} \frac{dP}{dx} = - \frac{\delta^*}{u_\tau^2} U_e \frac{dU_e}{dx} \label{beta}
\end{eqnarray}
and $f\bigg|_a^b = f(b) - f(a)$. Using the boundary conditions,
\begin{eqnarray}
 U\bigg|_a  =  0,\quad  U\bigg|_{a + k\delta}  =  U_e, \\
 V\bigg|_a  =  0,\quad  V\bigg|_{a + k\delta}  = V_o, \\
 \frac {\partial U}{\partial r}\bigg|_a  =  u_\tau^2/\nu, \quad \frac {\partial U} {\partial r}\bigg|_{a + k\delta} & = & 0, \\
 (-\overline{u'v'})\bigg|_a  = (-\overline{u'v'})\bigg|_{a + k\delta}  =  0,
\end{eqnarray}
the right hand side of eq. \ref{eq:ibl2} can be evaluated. This yields,
\begin{eqnarray}
\int_{a}^{a + k\delta} rU\frac{\partial U}{\partial x} dr + \int_{a}^{a + k\delta}  rV\frac{\partial U}{\partial r} dr  =  -\beta_{_{RC}}\frac{u_\tau^2}{2\delta^*}r^2\bigg|_a^{a + k\delta} -au_\tau^2 \nonumber \\
\implies \int_{a}^{a + k\delta} rU\frac{\partial U}{\partial x} dr + (rVU)\bigg|_a^{a + k\delta} - \int_{a}^{a + k\delta}  U\frac{\partial (rV)}{\partial r}dr = -\beta_{_{RC}}\frac{u_\tau^2}{2\delta^*}r^2\bigg|_a^{a + k\delta}  -au_\tau^2 \nonumber\\
\implies \int_{a}^{a + k\delta} rU\frac{\partial U}{\partial x} dr + (a + k\delta)V_o U_e +\int_{a}^{a + k\delta} rU\frac{\partial U}{\partial x}dr  = -\beta_{_{RC}}\frac{u_\tau^2}{2\delta^*}r^2\bigg|_a^{a + k\delta}  -au_\tau^2 \nonumber\\ 
\implies \int_{a}^{a + k\delta} r\frac{\partial U^2}{\partial x} dr = - (a + k\delta)V_o U_e  -\beta_{_{RC}}\frac{u_\tau^2}{2\delta^*}r^2\bigg|_a^{a + k\delta}  -au_\tau^2. \nonumber \\
\label{eq:b}
\end{eqnarray}

The shape factor, $H$ is defined as, 
\begin{eqnarray}
 H  & = &  \frac {\delta^*}{\theta}. \label{eq:hdef}
 \end{eqnarray}
 
 Differentiating both sides with respect to $x$,
 \begin{eqnarray}
\frac{dH}{dx} & = & \frac{1}{\theta}\frac{d\delta^*}{dx} - \frac{\delta^*}{\theta^2}\frac{d\theta}{dx} \\
\implies \theta \frac{dH}{dx} & = &  \frac{d\delta^*}{dx} - H\frac{d\theta}{dx} \\
\implies H & = & \frac {\frac {d\delta^*}{dx}}{\frac {d\theta}{dx}}  - \theta \frac {\frac{dH}{dx}}{\frac {d\theta}{dx}}.
\label{eq:h}
\end{eqnarray}

Note that no assumption has been made regarding the self-similarity of the boundary layer as yet. The second term in the right hand side of eq. \ref{eq:h} is small as $H$ varies very slowly with $x$ as compared to $\delta^*$ and hence, can be neglected. Self-similarity implies $\frac{dH}{dx}=0$, which makes the second term identically zero. Therefore, 
\begin{eqnarray}
H = \bigg(\frac {d\delta^*}{dx}\bigg) \bigg/ \bigg(\frac {d\theta}{dx}\bigg). \label{eq:hnew}
\end{eqnarray}

$\delta^*$ and $\theta$ for axisymmetric boundary layers are defined \citep{luxton1984} such that, 
\begin{eqnarray}
\label{eq:t1}
(\delta^* + a)^2 - a^2 & = & 2 \int_a^{a + \delta} \bigg(1- \frac {U}{U_e}\bigg)rdr, \\
(\theta + a)^2 - a^2 & = & 2 \int_a^{a + \delta} \frac{U}{U_e}\bigg(1- \frac {U}{U_e}\bigg)rdr. 
\label{eq:t2}
\end{eqnarray}
Note that $U = U_e$ for $r \ge \delta$, hence eqs. \ref{eq:t1} and \ref{eq:t2} can be written as,
\begin{eqnarray}
\label{eq:t1modified}
(\delta^* + a)^2 - a^2 & = & 2 \int_a^{a + k\delta} \bigg(1- \frac {U}{U_e}\bigg)rdr, \\
(\theta + a)^2 - a^2 & = & 2 \int_a^{a + k\delta} \frac{U}{U_e}\bigg(1- \frac {U}{U_e}\bigg)rdr, 
\label{eq:t2modified}
\end{eqnarray}
since $k \ge 1$.

Differentiating both sides with respect to $x$ and using the Leibniz integral rule in the right hand side yield,
\begin{eqnarray}
\label{eq:t3}
2(\delta^* + a)\frac{d\delta^*}{dx}  =  -\frac{2}{U_e} \int_a^{a + k\delta} \frac {\partial (rU)} {\partial x} dr + \frac{2}{U_e}\frac {dU_e} {dx} I \\
2(\theta + a)\frac{d\theta}{dx}  =  \frac{2}{U_e} \int_a^{a + k\delta} \frac {\partial (rU)} {\partial x} dr - \frac{2}{U_e} \frac {dU_e}{dx} I \nonumber \\
   -\frac{2}{U_e^2} \int_a^{a + k\delta} \frac {\partial (rU^2)}{\partial x}dr + \frac{4}{U_e}\frac {dU_e}{dx} J  \label{eq:t4}
\end{eqnarray}
where, 
\begin{eqnarray}
I & = & \int_a^{a + k\delta}\frac{U}{U_e}rdr,\\
J & = & \int_a^{a + k\delta}\frac{U^2}{U_e^2}rdr.
\end{eqnarray}

Using eqs. \ref{eq:ibl1} and \ref{eq:b} in the right hand side of eqs. \ref{eq:t3} and \ref{eq:t4} yield,
\begin{eqnarray}
\label{eq:t5}
2(\delta^* + a)\frac{d\delta^*}{dx}  =  2\frac{V_o}{U_e} (a + k\delta) -2\frac{\beta_{_{RC}}}{\delta^*}\frac{u_\tau^2}{U_e^2} I, \\
2(\theta + a)\frac{d\theta}{dx}  =  -2\frac{V_o}{U_e} (a + k\delta) +2\frac{\beta_{_{RC}}}{\delta^*}\frac{u_\tau^2}{U_e^2} I \nonumber \\
+2\frac{V_o}{U_e}(a + k\delta)  + \frac{\beta_{_{RC}}}{\delta^*} \frac{u_\tau^2}{U_e^2} r^2\bigg|_a^{a + k\delta}  + 2a\frac{u_\tau^2}{U_e^2} - 4 \frac{\beta_{_{RC}}}{\delta^*}\frac{u_\tau^2}{U_e^2} J \nonumber \\
\implies 2(\theta + a)\frac{d\theta}{dx}  =  2a\frac{u_\tau^2}{U_e^2} + 2\frac{\beta_{_{RC}}}{\delta^*}\frac{u_\tau^2}{U_e^2} I + \frac{\beta_{_{RC}}}{\delta^*} \frac{u_\tau^2}{U_e^2} r^2\bigg|_a^{a + k\delta} 
- 4\frac{\beta_{_{RC}}}{\delta^*}\frac{u_\tau^2}{U_e^2} J.
\label{eq:t6}
\end{eqnarray}

Dividing eq. \ref{eq:t5} by eq. \ref{eq:t6} and using eq. \ref{eq:hnew}  followed by rearranging the terms, we get,
\begin{eqnarray}
\label{eq:t7}
\bigg(\frac{\delta^* + a}{\theta + a} \bigg)H & = & \Bigg[ \frac {2\frac{V_o}{U_e} (a + k\delta) -2\frac{\beta_{_{RC}}}{\delta^*}\frac{u_\tau^2}{U_e^2} I}{2a\frac{u_\tau^2}{U_e^2} + 
2\frac{\beta_{_{RC}}}{\delta^*}\frac{u_\tau^2}{U_e^2} I + \frac{\beta_{_{RC}}}{\delta^*} \frac{u_\tau^2}{U_e^2} r^2\bigg|_a^{a + k\delta} - 4\frac{\beta_{_{RC}}}{\delta^*}\frac{u_\tau^2}{U_e^2} J}\Bigg ].  
\end{eqnarray}

Using the definitions of $\delta^*$ (eq. \ref{eq:t1modified}) and $\theta$ (eq. \ref{eq:t2modified}), it can be shown that,
\begin{eqnarray}
I & = & \frac{r^2}{2}\bigg|_a^{a + k\delta} - \frac{r^2}{2}\bigg|_a^{a + \delta^*} , \\
J & = & \frac{r^2}{2}\bigg|_a^{a + k\delta} - \frac{r^2}{2}\bigg|_a^{a + \delta^*} - \frac{r^2}{2}\bigg|_a^{a + \theta}.
\end{eqnarray}
Also, eq. \ref{eq:t5} yields,
\begin{eqnarray}
(\delta^* + a)\frac{d\delta^*}{dx}  & = &  \frac{V_o}{U_e} (a + k\delta) -\frac{\beta_{_{RC}}}{\delta^*}\frac{u_\tau^2}{U_e^2}I \nonumber \\
\implies \frac{V_o}{U_e} (a + k\delta) & = & (\delta^* + a)\frac{d\delta^*}{dx} + \frac{\beta_{_{RC}}}{2\delta^*}\frac{u_\tau^2}{U_e^2}\bigg(r^2\bigg|_a^{a + k\delta} - r^2\bigg|_a^{a + \delta^*} \bigg).
\label{eq:va}
\end{eqnarray}
Hence, eq. \ref{eq:t7} can be rearranged to show that,
 
 \begin{eqnarray}
 \bigg(\frac{\delta^* + a}{\theta + a} \bigg)H \bigg [2a\frac{u_\tau^2}{U_e^2} + \frac{\beta_{_{RC}}}{\delta^*}\frac{u_\tau^2}{U_e^2}\bigg(r^2\bigg|_a^{a + k\delta} - r^2\bigg|_a^{a + \delta^*}\bigg) + \frac{\beta_{_{RC}}}{\delta^*} \frac{u_\tau^2}{U_e^2} r^2\bigg|_a^{a + k\delta} \nonumber \\
- 2\frac{\beta_{_{RC}}}{\delta^*}\frac{u_\tau^2}{U_e^2}\bigg(r^2\bigg|_a^{a + k\delta} - r^2\bigg|_a^{a + \delta^*} - r^2\bigg|_a^{a + \theta}\bigg) \bigg ] & = & \\
2\frac{V_o}{U_e} (a + k\delta) - \frac{\beta_{_{RC}}}{\delta^*}\frac{u_\tau^2}{U_e^2} \bigg(r^2\bigg|_a^{a + k\delta} - r^2\bigg|_a^{a + \delta^*} \bigg) \nonumber \\
\implies 2\frac{V_oU_e}{u_\tau^2} (a + k\delta)\bigg (\frac{\theta + a}{\delta^* + a}\bigg) = H \bigg[2a + \frac{\beta_{_{RC}}}{\delta^*}\bigg(r^2\bigg|_a^{a + \delta^*} + 2r^2\bigg|_a^{a + \theta}\bigg) \bigg] \nonumber \\
+ \bigg (\frac{\theta + a}{\delta^* + a}\bigg) \frac{\beta_{_{RC}}}{\delta^*}\bigg(r^2\bigg|_a^{a + k\delta} - r^2\bigg|_a^{a + \delta^*}\bigg). \label{eq:heqn}
 \end{eqnarray}
Substituting for $V_o$ from eq. \ref{eq:va} and rearranging,
\begin{eqnarray}
 (\theta + a)\frac{d\delta^*}{dx} & = & H \frac{u_\tau^2}{U_e^2} \bigg[a + \frac{\beta_{_{RC}}}{2\delta^*}\bigg(r^2\bigg|_a^{a + \delta^*} + 2r^2\bigg|_a^{a + \theta}\bigg) \bigg] \nonumber \\
 \implies \frac{u_\tau^2}{U_e^2} & = & \frac {(\theta + a)\frac{d\delta^*}{dx}}{H\bigg[a + \frac{\beta_{_{RC}}}{2\delta^*}\bigg(r^2\bigg|_a^{a + \delta^*} + 2r^2\bigg|_a^{a + \theta}\bigg)\bigg ]} \nonumber \\
 \implies C_f & = &  \frac {2(1 + \frac{\theta}{a}) \frac{d\delta^*}{dx}}{H + \beta_{_{RC}}\bigg[2 + H\bigg(1 + \frac{\delta^*}{2a} + \frac{\theta^2}{a\delta^*} \bigg)\bigg]}.
\end{eqnarray}
Self-similarity of boundary layers implies that $\delta^*/\delta$ is constant.  So $C_f$ can be written as,
\begin{eqnarray}
 C_f = \frac {2(1 + \frac{\theta}{a}) \frac{\delta^*}{\delta}\frac{d\delta}{dx}}{H + \beta_{_{RC}}\bigg[2 + H\bigg(1 + \frac{\delta^*}{2a} + \frac{\theta^2}{a\delta^*} \bigg)\bigg]}. \label{r1}
\end{eqnarray}
Note that $C_f=2u_\tau^2/U_e^2$ is related to $\beta_{_{RC}}$ by definition (see eq. \ref{beta}). But that definition contains external flow parameters. On the other hand, eq. \ref{r1} relates $C_f$ to the boundary layer parameters directly. Also, eq. \ref{eq:heqn} can be rearranged to show that,
\begin{multline}
\frac{U_eV_o}{u_\tau^2}\bigg(\frac{1 + \theta/a}{1 + \delta^*/a}\bigg)\bigg(1 + k\frac{\delta}{a} \bigg) =  \\ H + \beta_{_{RC}} \bigg[2 + H\bigg(1 + \frac{\delta^*}{2a} + \frac{\theta^2}{a\delta^*}\bigg) + \bigg(\frac{1 + \theta/a}{1 + \delta^*/a}
\bigg) \bigg(k\frac{\delta}{\delta^*} -1 + \frac{k^2\delta^2 - \delta^{*2}}{2a\delta^*}\bigg) \bigg]\label{r2}
\end{multline}

At the edge of the boundary layer, $k=1$ and $V_o=V_e$. Therefore,
\begin{multline}
\frac{U_eV_e}{u_\tau^2}\bigg(\frac{1 + \theta/a}{1 + \delta^*/a}\bigg)\bigg(1 + \frac{\delta}{a} \bigg) =  \\ 
H + \beta_{_{RC}} \bigg[2 + H\bigg(1 + \frac{\delta^*}{2a} + \frac{\theta^2}{a\delta^*}\bigg) + \bigg(\frac{1 + \theta/a}{1 + \delta^*/a}
\bigg) \bigg(\frac{\delta}{\delta^*} -1 + \frac{\delta^2 - \delta^{*2}}{2a\delta^*}\bigg) \bigg]. \label{r3}
\end{multline}

At the verge of separation, $u_\tau$ goes to zero. Using the definition of $\beta_{_{RC}}$ (eq. \ref{beta}), eq. \ref{r3} yields,
\begin{multline}
V_e = -\delta^* \frac{dU_e}{dx} \bigg(\frac{1 + \theta/a}{1 + \delta^*/a}\bigg)^{-1}\bigg(1 + \frac{\delta}{a} \bigg)^{-1} \bigg[2 + H\bigg(1 + \frac{\delta^*}{2a} + \frac{\theta^2}{a\delta^*}\bigg) \\ 
+ \bigg(\frac{1 + \theta/a}{1 + \delta^*/a} \bigg) \bigg(\frac{\delta}{\delta^*} -1 + \frac{\delta^2 - \delta^{*2}}{2a\delta^*}\bigg) \bigg]. \label{vsep}
\end{multline}

\section{Comparison to previous work}
\label{sec:valid}
\subsection{Consistency with planar boundary layers relations}
For a planar boundary layer, $1/a$ approaches $0$ as $a$ approaches $\infty$. Setting $1/a=0$ in eq. \ref{r1} and \ref{r2} yields,
\begin{eqnarray}
\label{APG}
 C_f = \frac {2\frac{\delta^*}{\delta}\frac{d\delta}{dx}}{H + \beta_{_{RC}}\bigg(2 + H \bigg)}, \quad \text{and}\\
\frac{U_eV_o}{u_\tau^2} =  H + \beta_{_{RC}} \bigg(1 + H + k\frac{\delta}{\delta^*} \bigg). \label{APG_H} 
\end{eqnarray}

At the verge of separation, $u_\tau=0$; setting $k=1$ yields,
\begin{eqnarray}
 V_e =  -\delta^* \frac{dU_e}{dx} \bigg(1 + H + \frac{\delta}{\delta^*} \bigg).
\end{eqnarray}

These relations are identical to those derived by \cite{wei2017} (eq. 13 and 14 of their paper) for planar boundary layer with pressure gradient. They compared their analytical relations to the data available in literature for APG TBL and found good agreement (see figure 2-5 of their paper). 

Setting $\beta_{_{RC}} = 0$ in eq. \ref{APG_H} yields,
\begin{eqnarray}
 \frac {U_e V_o}{u_\tau^2} = H. \label{here}
\end{eqnarray}
Note that for $\beta_{_{RC}}=0$, regardless of the value of $k$, $V_o$ is same i.e. $V_o=V_e$ is constant outside the boundary layer. Eq. \ref{here} was derived by \cite{wei2016} (eq. 11 of their paper) and shown to be valid for laminar, transitional and turbulent boundary layers.

\subsection{Axisymmetric ZPG laminar boundary layer}
\begin{figure}
\begin{center}
\includegraphics[height=55mm]{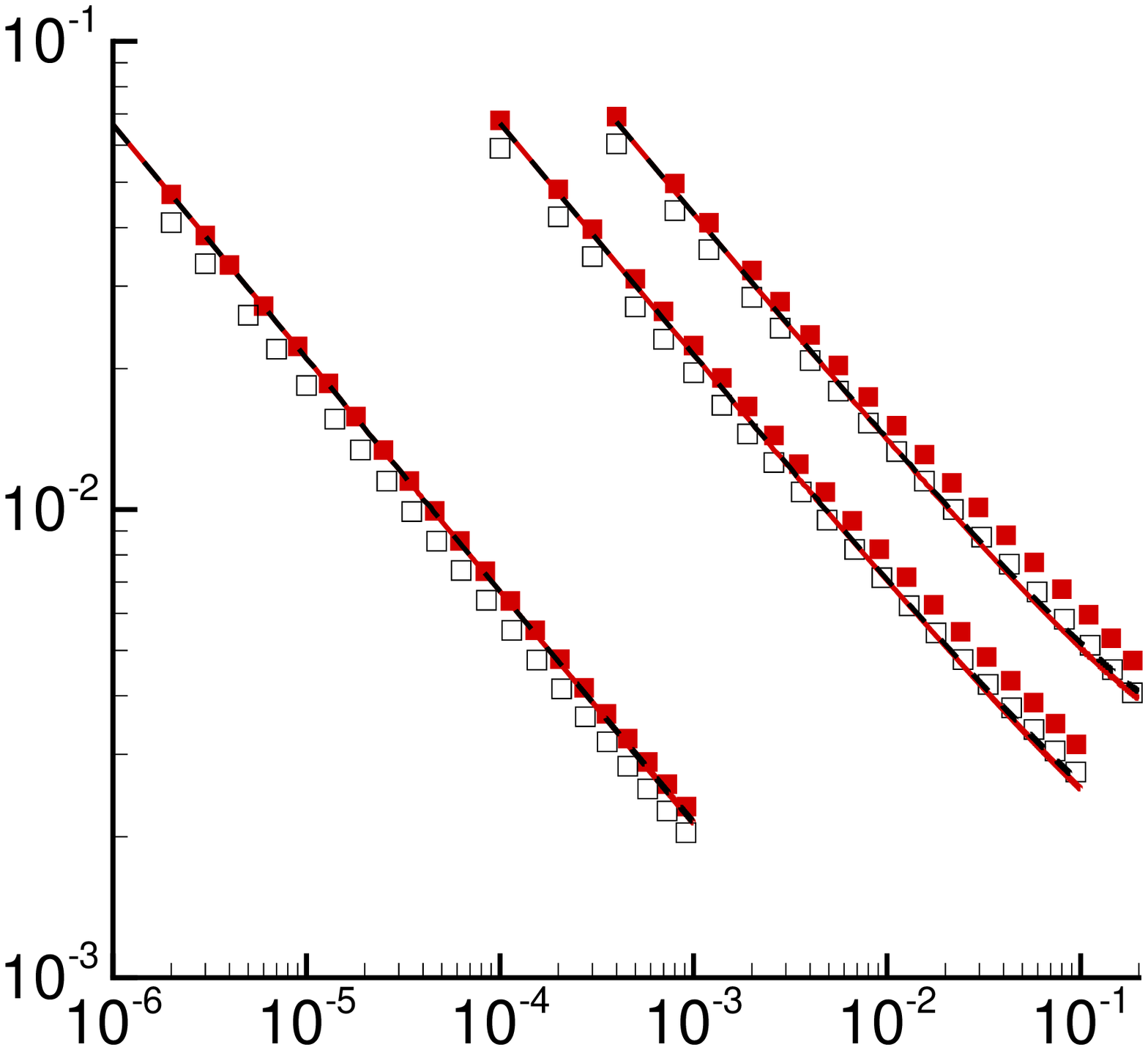}
\put(-100,0){$\frac{\nu x}{Ua^2}$}
\put(-180,80){$C_f$}
\put(-180,135){$(a)$}
\put(-32,32){$Re_a=1000$}
\put(-100,28){$Re_a=10000$}
\put(-45,90){$Re_a=500$} \hspace{8mm}
\includegraphics[height=55mm]{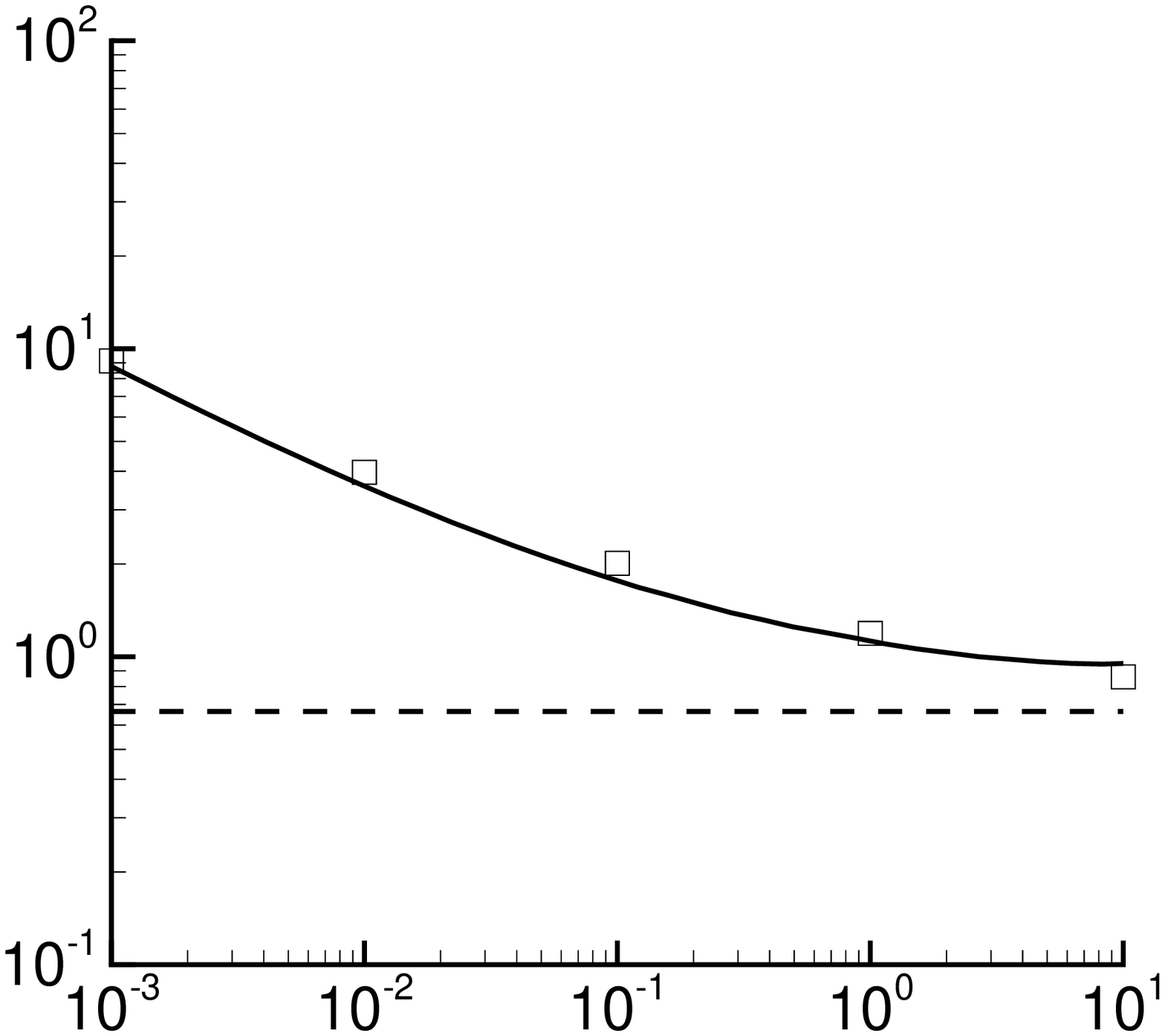}
\put(-100,0){$Re_a$}
\put(-200,80){$C_f \sqrt{Re_x}$}
\put(-180,135){$(b)$}
\end{center}
\caption{Skin-friction coefficient ($C_f$) as a function of non-dimensional parameter $\frac{\nu x}{Ua^2}$ (a), where results for radius based Reynolds number $Re_a=500$, 1000 and 10000 are shown along with solutions of Seban-Bond-Kelly \citep{seban,kelly}({\color{red} $\blacksquare$}) and Glauert--Lighthill \citep{gl}($\Box$). The present result (eq. \ref{zpglbl}) using $\delta^*$ from SBK ({\color{red}$-$}) and GL ($- -$), show identical $C_f$. $C_f$ as a function of $Re_a$ is compared with the result of \cite{cebeci1970} ($\Box$) for long thin cylinder (large $x/a$), where boundary layer thickness reaches asymptotic value \citep{stewartson}(b). The value obtained from the Blasius solution ($- -$) is also shown in (b) for comparison.}
\label{fig:lam}
\end{figure}

The SBK solution \citep{seban,kelly} for axisymmetric laminar boundary layer is valid up to  $\frac{\nu x}{Ua^2} < 0.04$, and was subsequently extended by \cite{gl} to the interval $0.04 < \frac{\nu x}{Ua^2} < 100$. For ZPG laminar axisymmetric boundary layer, eq. \ref{r1} becomes,
\begin{eqnarray}
C_{f, axisymmetric} = 2\frac{d\theta}{dx}\bigg(1 + \frac{\theta}{a} \bigg) = C_{f, planar} \bigg(1 + \frac{\delta^*}{aH} \bigg). \label{zpglbl}
\end{eqnarray}
 $\delta^*$ can be obtained from either SBK or GL solutions and $H=2.59$ for a laminar boundary layer. Thus, $C_f$ can be obtained. Figure \ref{fig:lam}(a) shows $C_f$ as a function of  $\frac{\nu x}{Ua^2}$ for three different $Re_a= 10000$, 1000 and 500, compared with both SBK and GL solutions. Note that the difference in $C_f$ using $\delta^*$ from either solution (SBK or GL) is negligible. Our results smoothly transitions from SBK to GL solution as $\frac{\nu x}{Ua^2}$ increases, as evident in the lower $Re_a$ cases. Figure \ref{fig:lam}(b) compares our result with the numerical solution of \cite{cebeci1970}, where $Re_a$ is varied. $\delta^*$ and $H$ for this case are estimated from the asymptotic results of \cite{stewartson}. The $C_f$ obtained from the Blasius solution ($C_f \sqrt{Re_x} = 0.664$) \citep{schlichting} is also shown for comparison. Overall, our results show good agreement with \cite{cebeci1970} for the entire range from thin to thick axisymmetric laminar boundary layer. Note that at large $Re_a$, $\delta/a$ approaches zero and hence, the axisymmetric laminar boundary layer approaches planar behaviour.   
 
\begin{figure}
\begin{center}
\includegraphics[height=55mm]{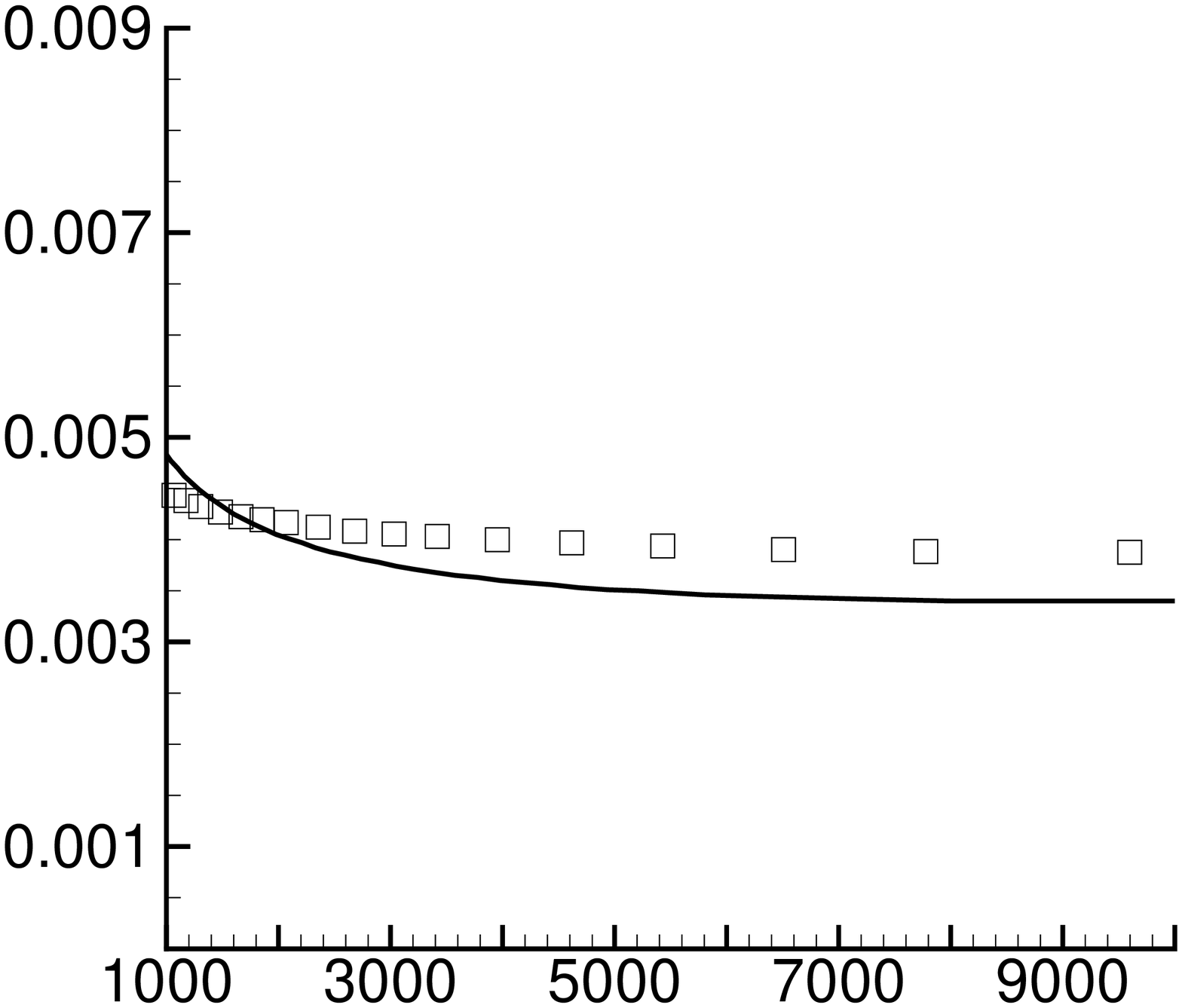}
\put(-100,0){$Re_\theta$}
\put(-190,135){$(a)$}
\put(-190,80){$C_f$}
\includegraphics[height=55mm]{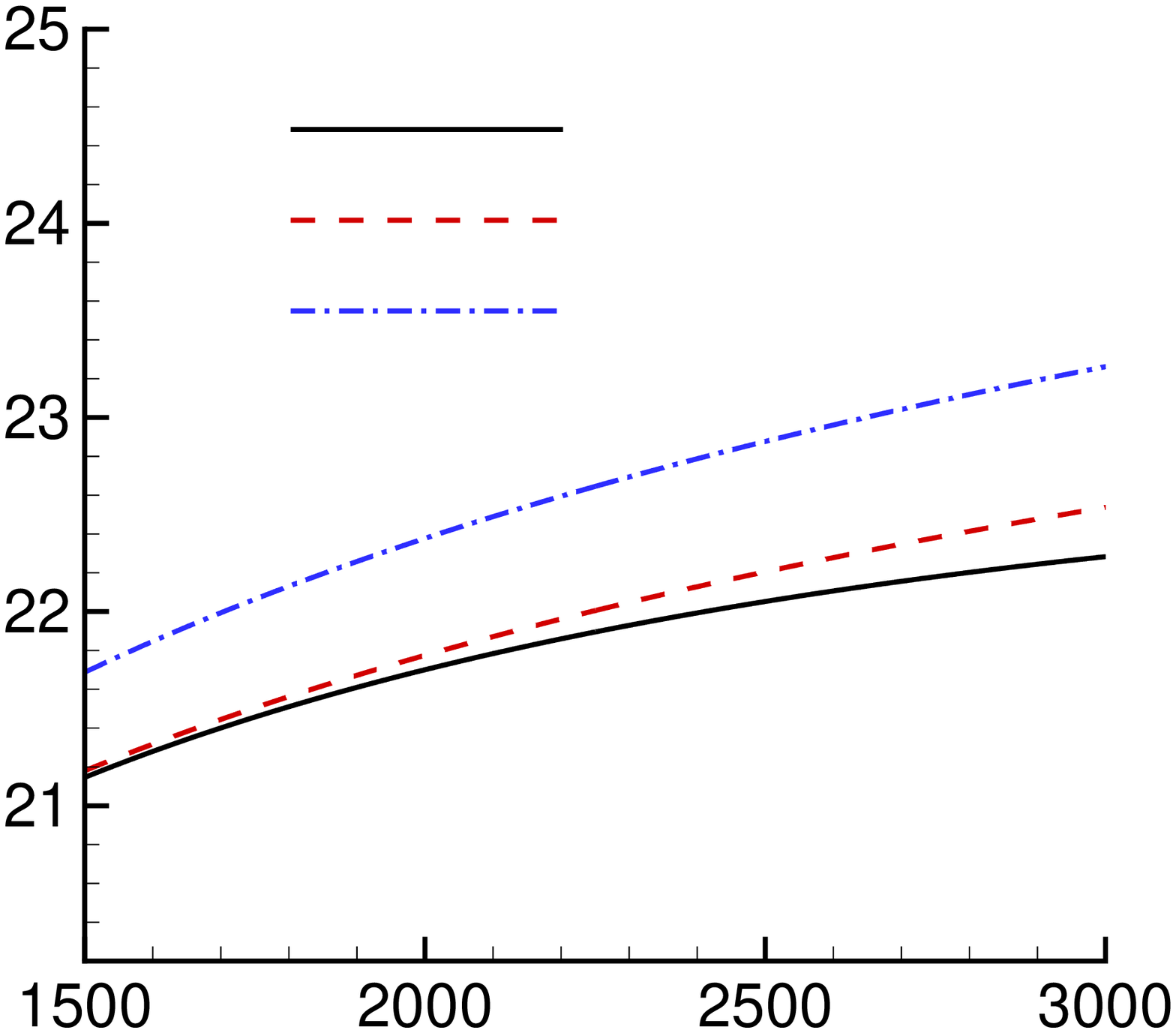}
\put(-100,0){$Re_\theta$}
\put(-78,122){present}
\put(-78,110){Monte et al.}
\put(-78,98){Woods}
\put(-185,135){$(b)$}
\put(-180,80){$U_e^+$}
\end{center}
\caption{(a) Skin-friction coefficient ($C_f$) as a function of $Re_\theta$ is compared with the result of \cite{cebeci1970}($\Box$) for slender cylinder for radius based Reynolds number $Re_a=40200$ and radius $a=1''$. The shape factor of $H=1.4$ and $C_{f,planar}$ correlation of \cite{monkewitz2008} is used in our relation to predict $C_f$. The boundary layer growth is assumed identical to that of flat plate, which need not be true for slender cylinders at high $Re_\theta$. (b) $U_e^+$ as a function of $Re_\theta$ is compared with the correlations of \cite{monte2011} and \cite{woods}. $U_e^+$ is related to $C_f$ as $U_e^+ = \sqrt{2/C_f}$ .}
\label{fig:turb}
\end{figure}

\subsection{Axisymmetric ZPG turbulent boundary layer}
\cite{cebeci1970} numerically solved incompressible turbulent ZPG axial flow over a circular slender cylinder of radius, $a=1''$ and $Re_a=40200$. The same relation eq. \ref{zpglbl} is used to estimate $C_f$ but the $C_{f,planar}$ correlation of \cite{monkewitz2008} is used. The shape factor $H$ is assumed to be 1.4 and the boundary layer growth $d\theta/dx$ is assumed identical in both planar and axisymmetric case. Figure \ref{fig:turb} (a) shows our results compared to that of \cite{cebeci1970}. Note that the range of $Re_\theta$ on the cylinder is large ($1000 < Re_\theta <10000$). Hence, the assumption of identical growth and $H=1.4$ may not hold, which is the reason for the difference between our result and that of \cite{cebeci1970}. In reality, $H$ is a weakly decreasing function of $Re_\theta$ for TBL \citep{monkewitz2008}. For example, $H \approx 1.45$ at $Re_{\theta} = 1000$ \citep{schlatter2010}, whereas $H \approx 1.36$ at $Re_{\theta} = 9000$ \citep{osterlund}. The results shown in figure \ref{fig:turb} (a) will further improve if the variation of $H$ with $Re_{\theta}$ is taken into account.      

\cite{kumar_snh16} simulated thin axisymmetric TBL in the range $1400 < Re_\theta < 1620$. Using their boundary layer $\delta^*$ and $\theta$ variation with streamwise distance $x$, which is almost linear, their slope $d\delta^*/dx$ and $d\theta/dx$ can be estimated. This estimated slope can be used to compute $C_f$ for $1500 < Re_\theta < 3000$ as shown in figure \ref{fig:turb}(b). Our results are compared with correlation of \cite{monte2011}, which corrected the correlation of \cite{woods} using their extensive simulation database, showing good agreement. Note that for a large range of $Re_\theta$, the assumption of linear growth of boundary layer breaks down, hence the differences at large $Re_\theta$. 
  
\section{Discussion}
\label{sec:dis}

\subsection{Effect of curvature on $C_f$}
If both planar and axisymmetric boundary layers have the same boundary layer parameters, eqs. \ref{r1} and \ref{APG} yield: 
\begin{eqnarray}
\label{eq:eta}
 \frac {C_{f, axisymmetric}}{C_{f, planar}}  =  \frac{\bigg(1 + \frac{\theta}{a}\bigg) \bigg[H + \beta_{_{RC}}\bigg(2 + H \bigg)\bigg]}{H + \beta_{_{RC}}\bigg[2 + H\bigg(1 + \frac{\delta^*}{2a} + \frac{\theta^2}{a\delta^*} \bigg)\bigg]} \nonumber \\
 \implies \frac{C_{f, axisymmetric}}{C_{f, planar}} - 1  
= \frac {\frac{\theta}{a}H + \beta_{_{RC}} \bigg[\frac{\theta}{a} + \frac{\delta^*}{a} \bigg(1 - \frac{H}{2} \bigg) \bigg]} {H + \beta_{_{RC}}\bigg[2 + H\bigg(1 + \frac{\delta^*}{2a} + \frac{\theta^2}{a\delta^*} \bigg)\bigg]}.
\label{finalratio}\end{eqnarray}

Thus, if the right hand side of eq. \ref{finalratio} is positive, the presence of curvature increases $C_f$ and vice-versa. 

It is easy to see that for ZPG ($\beta_{_{RC}}=0$) boundary layers, 
\begin{eqnarray}\label{zero}
\frac {C_{f, axisymmetric}}{C_{f, planar}} = 1 + \frac{\theta}{a}.
\end{eqnarray}
For boundary layer with APG ($\beta_{_{RC}} > 0$), the denominator of the right hand side of eq. \ref{finalratio} is always positive. Hence, the effect of curvature will depend on the sign of the numerator $\eta$ defined as, 
\begin{eqnarray}
 \eta = \frac{\theta}{a}H + \beta_{_{RC}} \bigg[\frac{\theta}{a} + \frac{\delta^*}{a} \bigg(1 - \frac{H}{2} \bigg) \bigg].
\end{eqnarray}
It can be shown that $\eta \ge 0$ if $\beta_{_{RC}} \ge 0$ (see appendix \ref{A1}). Therefore, the presence of curvature increases $C_f$ if $\beta_{_{RC}} \ge 0$. Note that, this is true regardless of the value of $a$. It has been assumed that $d\delta/dx$ is identical for both planar and axisymmetric TBL. This is not be always true. In fact, for thick axisymmetric TBL at zero-pressure-gradient ($\delta/a \gg 1$ and $\beta_{_{RC}}=0$), $d\delta/dx$ is smaller than that of planar TBL value \citep{tutty}. However, $C_f$ is still higher than planar values because $\theta/a \gg 1$, which compensates for the decrease in $d\delta/dx$. 

The presence of curvature may or may not increase $C_f$ in FPG axisymmetric TBL depending on the sign of the right hand side of eq. \ref{finalratio}.  

\subsection{Thick axisymmetric ZPG turbulent boundary layer}
For $\beta_{_{RC}}=0$, the expression for $C_f$ (eq. \ref{r1}) reduces to,

\begin{eqnarray}
 C_f = 2\bigg(1 + \frac{\theta}{a}\bigg)\frac{\theta}{\delta} \frac{d\delta}{dx}. 
\end{eqnarray}
Thus, knowing local boundary layer parameters, $C_f$ can be estimated. For example, \cite{jordan2014jfe}  compiled numerous experimental results along with his simulation database for thick axisymmetric TBL in ZPG and showed that $\delta/\theta \approx 7.2$. The estimated value of $d\delta/dx \approx 2.5 \times 10^{-3}$ for a range of thick axisymmetric TBL ($2.1 \le \delta/a \le 11$, $37 \le a^+ \le 388$, $586 \le Re_a \le 7475$). This makes,

\begin{eqnarray}
 C_f = 6.94 \times 10^{-4} \bigg(1 + \frac{\theta}{a}\bigg) = 6.94 \times 10^{-4} \bigg(1 + \frac{Re_\theta}{Re_a}\bigg).
\end{eqnarray} 

\subsection{Axisymmetric turbulent boundary layer under large APG}
For large APG, $\beta_{_{RC}} \gg 1$. Thus eq. \ref{r1} yields,
\begin{equation}
 C_f \approx \Bigg[\frac {2(1 + \frac{\theta}{a}) \frac{\delta^*}{\delta}\frac{d\delta}{dx}}{2 + H\bigg(1 + \frac{\delta^*}{2a} + \frac{\theta^2}{a\delta^*} \bigg)}\Bigg] \frac{1}{\beta_{_{RC}}}.
\end{equation}
For self-similar TBL in APG, $\delta^*/\delta$, $H$ and $d\delta/dx$ become constant \citep{maciel2006}. Similar behaviour is expected for axisymmetric TBL as well. When $\delta/a < 1$, $\theta/a$ and $\delta^*/a$ are small as compared to 1. This makes, the term inside brackets ([ ]) nearly constant. Thus for thin axisymmetric TBL at large APG, $C_f \sim 1/\beta_{_{RC}}$. A similar result was obtained by \cite{wei2017} for planar TBL. 

\subsection{Axisymmetric turbulent boundary layer under FPG} 
For FPG TBL, there are two important flow parameters: pressure gradient parameter ($\Lambda$) \citep{narasimha1973} and acceleration parameter ($K$) \citep{launder1964} defined as,
\begin{eqnarray}
\Lambda &=& - \frac{\delta}{u_\tau^2}\frac{1}{\rho}\frac{dP}{dx}, \\
K &=& \frac{\nu}{U_e^2} \frac{dU_e}{dx}.
\end{eqnarray}

All the relations derived in \S \ref{sec:gov} holds for FPG axisymmetric TBL as well by replacing $\beta_{_{RC}}$ with $-\Lambda$. It can be shown that,
\begin{eqnarray}
\frac{dC_f}{d\Lambda} = C_f \Bigg[\frac{2 + H\bigg(1 + \frac{\delta^*}{a} + \frac{\theta^2}{a\delta^*}\bigg)} {H - \Lambda \bigg[2 + H\bigg(1 + \frac{\delta^*}{a} + \frac{\theta^2}{a\delta^*} \bigg) \bigg]} \Bigg]
< 0.
\end{eqnarray}
Thus, increasing FPG decreases $C_f$ and this effect is expected to be enhanced by the presence of transverse curvature as the presence of terms with $1/a$ enhance the magnitude of $dC_f/d\Lambda$.

\section{Conclusion}
\label{sec:sum}
In this work, the integral analysis of equations governing axisymmetric boundary layer flow is presented, including the effect of pressure gradient. Analytical relations are derived relating $C_f$ to the boundary layer parameters. The relations for planar TBL with and without pressure gradient presented by \cite{wei2017} and \cite{wei2016} respectively can be recovered by setting $1/a = 0$ and further setting $\beta_{_{RC}}=0$. It has been shown that the presence of transverse curvature increases $C_f$ regardless of the nature of boundary layer, consistent with the observations reported in the literature for both ZPG and APG axisymmetric boundary layers. The derived relations are compared to the existing results in the literature showing good agreement. The results presented in this work are expected to be valid for any boundary layer as long as the governing equations hold, which assumes local dynamic equilibrium. It is challenging, both experimentally and computationally, to obtain accurate $C_f$ at high $Re$. However, it is relatively easier to obtain accurate mean velocity profiles. In addition to predicting the influence of pressure gradient and curvature, the derived expressions are potentially useful to both skin-friction measurements and wall-modelled large eddy simulation of turbulent boundary layers.      

\section*{Acknowledgement}
This work is supported by the United States Office of Naval Research (ONR) under ONR Grant N00014-14-1-0289 with Dr. Ki-Han Kim as technical monitor. We thank Mr. S. Anantharamu for useful discussions. 
\appendix
\section{Maximum value of $\eta$}
\label{A1}
It is known that, $H\ge 1$ which yields,
\begin{eqnarray}
 \frac{H}{2} \ge \frac{1}{2} \implies \frac{H}{2} - 1 \ge - \frac{1}{2}, \label{1} \\
 \frac{1}{H} \le 1 \implies -\frac{1}{H} \ge - 1. \label{2}
\end{eqnarray}
Adding eqs. \ref{1} and \ref{2} we get,
\begin{eqnarray}
 \frac{H}{2} - 1 -\frac{1}{H} \ge - \frac{3}{2}, \implies \frac{1}{\frac{H}{2} -1 - \frac{1}{H}} \le -\frac{2}{3}.\label{3}
 \end{eqnarray}
 But, 
 \begin{eqnarray}
 \frac{1}{\frac{H}{2} -1 - \frac{1}{H}} = \frac{H}{H(\frac{H}{2} -1) - 1} = \frac{-\frac{\theta}{a}H} {\frac{\theta}{a} + H\frac{\theta}{a}\bigg(1 - \frac{H}{2}\bigg)} = \frac{-\frac{\theta}{a}H} {\frac{\theta}{a} + \frac{\delta^*}{a}\bigg(1 - \frac{H}{2}\bigg)}.\label{4}
\end{eqnarray}
From eqs. \ref{3} and \ref{4}, it follows that,
\begin{eqnarray}
 \frac{-\frac{\theta}{a}H} {\frac{\theta}{a} + \frac{\delta^*}{a}\bigg(1 - \frac{H}{2}\bigg)} \le -\frac{2}{3}.\label{5}
\end{eqnarray}
Now,
\begin{eqnarray}
 \eta = \frac{\theta}{a}H + \beta_{_{RC}} \bigg[\frac{\theta}{a} + \frac{\delta^*}{a} \bigg(1 - \frac{H}{2} \bigg) \bigg] > 0 \nonumber \\
 \iff \beta_{_{RC}} > \frac{-\frac{\theta}{a}H} {\frac{\theta}{a} + \frac{\delta^*}{a}\bigg(1 - \frac{H}{2}\bigg)}. \label{6}
\end{eqnarray}
Using eq. \ref{5}, it is easy to see that eq. \ref{6} always holds for $\beta_{_{RC}} > 0$.
\bibliographystyle{jfm}
\bibliography{submit}
\end{document}